\documentclass{article}
\usepackage[T1]{fontenc}
\usepackage[utf8]{inputenc}
\usepackage{ismir} % Remove the "submission" option for camera-ready version
\usepackage{amsmath,cite,url}
\usepackage{graphicx}
\usepackage{color}
\usepackage{microtype}
\usepackage{enumitem}

\title{WeaveMuse: An Open Agentic System for Multimodal \\ Music Understanding and Generation}

\oneauthor
  {Emmanouil Karystinaios}
  {Institute of Computational Perception, Johannes Kepler University Linz, Austria \\\texttt{firstname.lastname@jku.at}}

\def\authorname{E. Karystinaios}

\begin{document}

\maketitle

\begin{abstract}
Agentic AI has been standardized in industry as a practical paradigm for coordinating specialized models and tools to solve complex multimodal tasks. In this work, we present WeaveMuse, a multi-agent system for music understanding, symbolic composition, and audio synthesis. Each specialist agent interprets user requests, derives machine-actionable requirements (modalities, formats, constraints), and validates its own outputs, while a manager agent selects and sequences tools, mediates user interaction, and maintains state across turns. The system is extendable and deployable either locally, using quantization and inference strategies to fit diverse hardware budgets, or via the HFApi to preserve free community access to open models. Beyond out-of-the-box use, the system emphasizes controllability and adaptation through constraint schemas, structured decoding, policy-based inference, and parameter-efficient adapters or distilled variants that tailor models to MIR tasks. A central design goal is to facilitate intermodal interaction across text, symbolic notation and visualization, and audio, enabling analysis-synthesis-render loops and addressing cross-format constraints. The framework aims to democratize, implement, and make accessible MIR tools by supporting interchangeable open-source models of various sizes, flexible memory management, and reproducible deployment paths.

\end{abstract}

% \textbf{Keywords:} deployment, quantization, autonomous WeaveMuses, prompt engineering, tool use, symbolic music, audio generation
\section{Introduction}
Large language models increasingly act as planners that coordinate specialized tools for music tasks spanning text, symbolic notation, and audio. At the same time, practical use is often limited by the cost of inference, the difficulty of deploying heterogeneous models as tools, and the lack of mechanisms for cross-modal control.

In this paper, we introduce WeaveMuse, an open, agentic system that orchestrates music understanding, generation, and analysis across modalities. A manager agent interprets user goals, maintains dialogue and task state, and composes pipelines by selecting among specialist agents and tools. Specialist agents translate requests into machine-actionable specifications (modalities, formats, constraints), execute analysis or generation, and perform basic self-checks against musical requirements. The system is designed to be extendable and modest in its assumptions: it can run locally by using quantization and other efficiency strategies to fit diverse hardware budgets, or be accessed via HFApi, so that open models remain freely accessible to the community.

% Rather than relying solely on out-of-the-box deployable models, WeaveMuse emphasizes controllability and adaptation. The runtime favors model plurality and interchangeability, using alternative open-source backbones of different sizes and agent/tool implementations that can be swapped according to resource budgets, while common logging records inference policies and resource use to support reproducibility.

We develop the WeaveMuse as a reference system and toolkit to lower the barrier to cross-modal interaction between text, score, and audio. Compared to other music agentic frameworks such as~\cite{yu2023musicagent, deng2024composerx, liu2024mumu}, WeaveMuse is positioned as an open, multi-agent, modular framework that focuses equally on both symbolic and audio representations. The system serves as a demonstration and adaptation of, for the moment a few, useful tools that the MIR community develops. The paper focuses on deployment and usability considerations, reports initial behaviors under constrained settings, and outlines limitations and directions for integrating, distilling, and extending models for finer control and more interactive music creation.

\section{System Overview}

\begin{figure*}[t]
    \centering
    \includegraphics[width=0.65\textwidth]{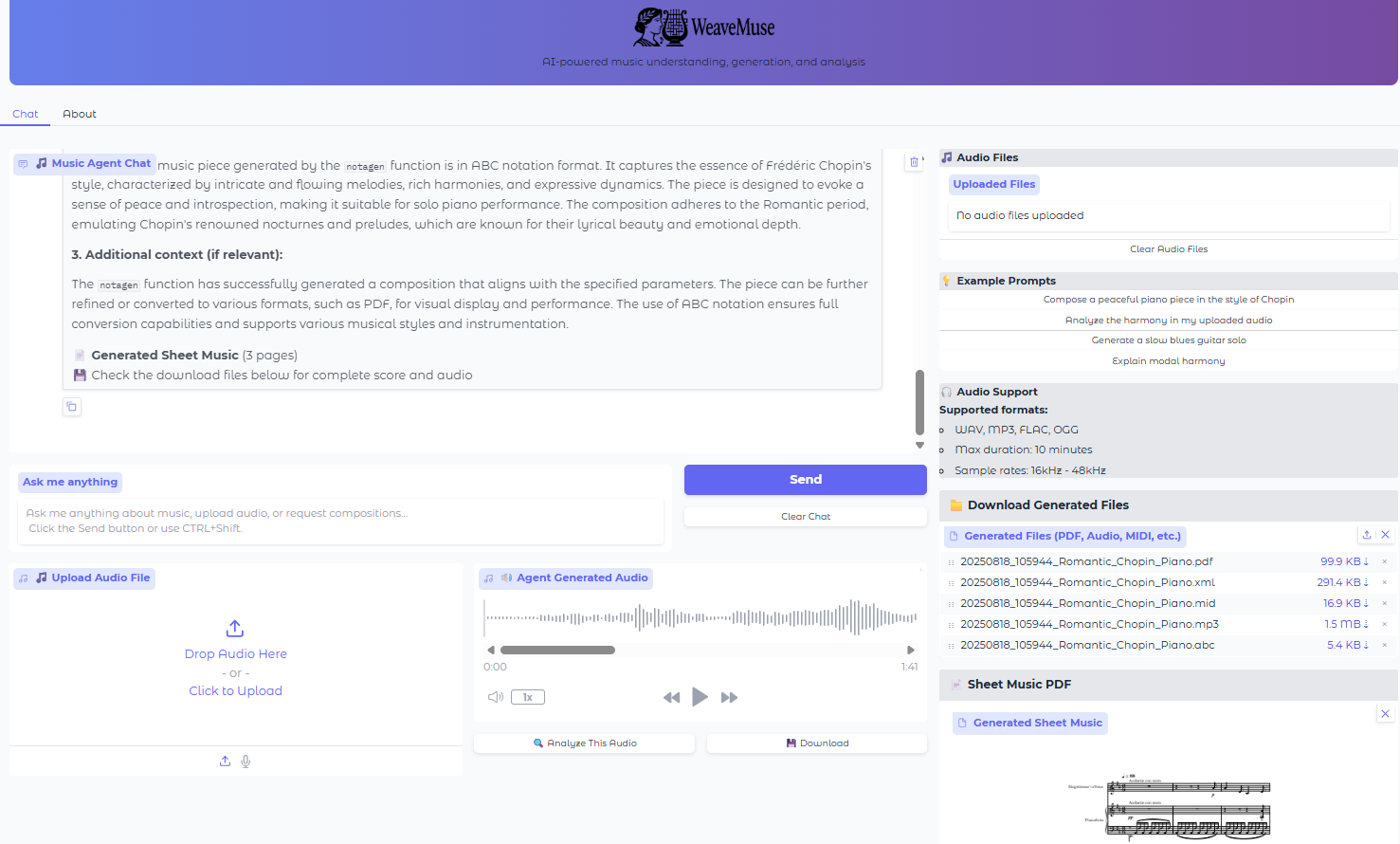}
    \caption{The WeaveMuse interface.}
    \label{fig:music-agent}
\end{figure*}

\subsection{Core agent} 
The WeaveMuse core agent maintains task context and a tool/specialized agent router that selects and sequences tools while considering resource hints. It is also responsible for applying reasoning or verification strategies. All agents are built with the smolagents library~\cite{smolagents}.

\subsection{Integrated tools} 
Adapters include: ChatMusician~\cite{yuan2024chatmusician} for music-theory reasoning and text/symbolic grounding; NotaGen~\cite{wang2025notagen} for ABC notation generation with pdf and audio compilation; Audio Analysis~\cite{goel2025audio, chu2023qwen} models for segmentation and understanding; Stable Audio Open~\cite{evans2025stable} for 44.1\,kHz stereo synthesis; and Score Visualization for sheet rendering with MuseScore and score engraving correction using~\cite{foscarin2024cluster}. Alternative open models/agents of varying sizes can be swapped per task and budget, supporting progressive enhancement (small/large when resources permit). Some adapters include their own agent, which is responsible for formatting arguments and prompts as well as adapting user queries.

\subsection{Interfaces and Deployment Modes} 

Both local and hosted interfaces are the identical. An example of the WeaveMuse interface is displayed in figure~\ref{fig:music-agent}. Interface is based on Gradio~\cite{abid2019gradio} with custom bindings and tools to integrade different modalities and their interactions. It offers to the user a straghtforward and modular GUI. Local interface implements models locally while the hosted inteface leverages HuggingFace Spaces and API for model calls.

\section{Autonomous Music Agents}
The agents in WeaveMuse follows a perceive, plan, act loop with verification and repair when possible:
\begin{enumerate}[noitemsep]    
    \item \textbf{Perceive:} Ingest text, audio, or symbolic inputs. 
    \item \textbf{Plan:} Compose a tool graph (e.g. enrich query, compose, engrave, synthesize), subject to explicit constraints and resource hints.
    \item \textbf{Act:} Invoke tools, use cache intermediates (ABC, MIDI, stems, analysis reports).
    \item \textbf{Critique \& repair:} When execution fails, or after a tool action is operated, analysis could be run to verify if the result matches the user's query.
\end{enumerate}

% \section{Inference \& Prompt Engineering}
% \subsection{Representation-aware prompting} 
% We reflect target modality: \emph{analysis prompts} (sections, keys, chords), \emph{composition prompts} (form, harmonic path, texture), \emph{conversion prompts} (ABC/MIDI bindings), and \emph{critique prompts} (rule checks). Prompts are parameterized by a constraint schema (key, tempo, meter, bars, style, instrument palette).

\subsection{Quantization \& Efficiency}
To operate under limited memory and compute while preserving musical quality, we employ quantization techniques and memory offloading. Furthermore, we support (i) dynamic precision switching per tool; (ii) CPU/GPU device placement and paging; (iii) lazy loading and on-disk caching for models and embeddings; and (iv) memory-aware batching for offline analysis jobs.

\subsection{Deployment Considerations}
\textbf{Local}: resource tiers (low/medium/high VRAM) map to default policies (INT4/INT8, cache offloading, attention kernels), ensuring functionality across different resource capacities. For reference, our high-resource tier capacity using \textrm{Qwen3-Coder-30B}~\cite{yang2025qwen3} as the agent and reasoning backbone is achieved on a single NVidia A40.
\textbf{Hosted}: HFApi access to open models provides immediate availability and community sharing. By using dynamic GPU allocation for the hosted application, we enable usability by everyone without overhead costs to the user or the host. The same planner and prompts run in both modes, aiding reproducibility.

\subsection{Limitations}
Agent-based systems are usually as efficient and effective as the underlying LLM model is potent. Computation or budget limitations can effectively downgrade the performance of the entire system. Furthermore, as this is a work in progress, tool orchestration and agentic prompting might not always work as expected. Smaller models (i.e. less than 3B parameters) do not always use the correct tools when interacting with user queries.

% Despite efficiency measures, (i) weight-only quantization can degrade subtle harmonic voice-leading in small models; (ii) cache quantization may introduce instability in very long sequences; (iii) audio rendering may not perfectly reflect symbolic microstructure; (iv) cross-model drift can occur when swapping backbones mid-session; and (v) constrained hardware may force smaller context windows, impacting large-form works.

\subsection{Availability \& Reproducibility}
The framework is open-source with a public repository.\footnote{Code: \url{github.com/manoskary/weavemuse}} Local and hosted configurations share identical planner and prompt templates. HFApi access to open models lowers the barrier for community evaluation.

\section{Conclusion \& Future Work}
The framework demonstrates that an efficiency-first, agentic stack can deliver controllable end-to-end pipelines under tight resource budgets. Future work will focus on the integration and distillation of additional open models (symbolic and audio), on the addition of more structured control over form/mixing, and on improving cross-format interaction (notation/text/audio) via alignment signals. We aim to release distilled checkpoints and adapter suites, plus automatic policy selection from hardware probes for seamless local or hosted deployment.

\section{Acknowledgments} 
This work was supported by the European Research Council (ERC) under Horizon 2020 grant \#101019375 “Whither Music?”.

% For BibTeX users:
\bibliography{ISMIR_LLM4Music_template}

% For non BibTeX users:
%\begin{thebibliography}{citations}
% \bibitem{Author:17}
% E.~Author and B.~Authour, ``The title of the conference paper,'' in {\em Proc.
% of the Int. Society for Music Information Retrieval Conf.}, (Suzhou, China),
% pp.~111--117, 2017.
%
% \bibitem{Someone:10}
% A.~Someone, B.~Someone, and C.~Someone, ``The title of the journal paper,''
%  {\em Journal of New Music Research}, vol.~A, pp.~111--222, September 2010.
%
% \bibitem{Person:20}
% O.~Person, {\em Title of the Book}.
% \newblock Montr\'{e}al, Canada: McGill-Queen's University Press, 2021.
%
% \bibitem{Person:09}
% F.~Person and S.~Person, ``Title of a chapter this book,'' in {\em A Book
% Containing Delightful Chapters} (A.~G. Editor, ed.), pp.~58--102, Tokyo,
% Japan: The Publisher, 2009.
%
%\end{thebibliography}

\end{document}